\input{aipcheck}

\documentclass[
    ,final            
  ]
  {aipproc}

\layoutstyle{6x9}

\newcommand{\be}{\begin{equation}}
\newcommand{\ee}{\end{equation}}
\newcommand{\bea}{\begin{eqnarray}}
\newcommand{\nn}{\nonumber}
\newcommand{\eea}{\end{eqnarray}}
\newcommand{\bn}{\mathbf \nabla}

\begin{document}

\title{Measuring mass moments and electromagnetic
moments of a massive, axisymmetric body, through gravitational
waves.}


\author{Theocharis A.~Apostolatos}{
  address={Section of Astrophysics, Astronomy, and Mechanics,
Department of Physics, National and Kapodistrian University of
Athens Panepistimiopolis, Zografos GR-15783, Athens, Greece.},
  email={tapostol@cc.uoa.gr},
}

\author{Thomas P.~Sotiriou}{
  address={SISSA, International School for Advanced Studies,
via Beirut 4, 34014 Trieste, Italy and \\INFN Sezione di Trieste.},
  email={sotiriou@sissa.it},
}

\begin{abstract}
 The electrovacuum around a rotating massive body with electric
charge density is described by its multipole moments (mass moments,
mass-current moments, electric moments, and magnetic moments). A
small uncharged test particle orbiting around such a body moves on
geodesics if gravitational radiation is ignored. The waves emitted
by the small body carry information about the geometry of the
central object, and hence, in principle, we can infer all its
multipole moments. Due to its axisymmetry the source is
characterized now by four families of scalar multipole moments: its
mass moments $M_l$, its mass-current moments $S_l$, its electrical
moments $E_l$ and its magnetic moments $H_l$, where
$l=0,1,2,\ldots$. Four measurable quantities, the energy emitted by
gravitational waves per logarithmic interval of frequency, the
precession of the periastron (assuming almost circular orbits), the
precession of the orbital plane (assuming almost equatorial orbits),
and the number of cycles emitted per logarithmic interval of
frequency, are presented as power series of the newtonian orbital
velocity of the test body. The power series coefficients are simple
polynomials of the various moments.
\end{abstract}

\maketitle


\section{Introduction}

The geometry around a massive object is uniquely characterized byo
the multipole moments \cite{Geroch,Hansen,Simon1,Simon2} of the
central object, as in newtonian gravity. Ryan \cite{Ryan} showed how
we could reveal the spacetime geometry of a central axisymmetric
body, by measuring a few physical quantities that are related to the
kinematics of a hypothetical test body that is orbiting around the
central object while it is emitting gravitational waves. This
geometry-mapping could be achieved in principle by an advanced
network of gravitational wave detectors with high sensitivity. Of
course,  to  determine all moments we need infinitely sensitive
detectors. However, knowing a few moments of the central body, we
could put restrictions on the various models that are assumed to
describe the interior of the central body.

Since the electromagnetic moments of the central object have their
own contribution on the geometry around the central object, we could
generalize the work of Ryan to yield the relation of all moments to
the physical quantities used by Ryan.

In our paper we have considered binaries consisting of (i) a central
object is assumed to be stationary and axisymmetric, and
characterized by reflection symmetry with respect to its equatorial
plane, (ii) a test particle with no charge that is orbiting around
the first one on a nearly circular and equatorial geodesic orbit.
The axisymmetry of the central object ensures that the spacetime
geometry could be described by scalar moments. Since we are dealing
with electrovaccuum solutions 4 sets of scalar moments are needed:
Its mass moments $M_0, M_2, M_4,\ldots$, its mass-current moments
$S_1,S_3,\ldots$, its electric charge moments $E_0,E_1,E_2,\ldots$,
and its magnetic moments $H_0,H_1,H_2,\ldots$, where e.g., the
$M_0\equiv M$ moment is the mass of the object, $S_1$ is its angular
momentum, $E_0\equiv E$ is its charge, and $H_{1}$ is its magnetic
dipole moment. In every set of moments, each moment appears in steps
of 2, and this holds good for the electromagnetic moments as well,
due to the reflection symmetry of the metric itself (cf.,
\cite{Hansen}). The fact that all electromagnetic moments are
written above, instead of every second one, is related to the fact
that two kind of objects could have axisymmetric metrics, either the
ones with even electric moments and odd magnetic moments, or the
other way around.

The two precessing frequencies of the orbits (the periastron
precession and the orbital precession) are written as functions of
the moments of the metric on which the particle is moving. Moreover,
although the orbit of the test body is assumed to be a geodesic one,
we are going to use the quadrupole formula to infer the energy
emitted by the binary. The evolution of the orbit will then be used
to obtain alternative information about the moments of the geometry.

Throughout the paper units are chosen so that $G=c=1$.

\section{The kinematics of the test particle}

If we perturb a circular equatorial orbit on a fixed axisymmetric
background then two new frequencies arise,
the one connected to the periastron precession and the one connected
to the orbital precession, that are related to the circular orbit
frequency and the metric itself by \cite{Ryan}

 \bea 
 \label{omegas}
\Omega_\alpha &=& \Omega- \left( -\frac{g^{\alpha\alpha}}{2} \left[
(g_{tt} + g_{t \phi} \Omega)^2
    \left(\frac{g_{\phi \phi}}{\rho^2} \right)_{,\alpha \alpha} \right. \right. {} \nn \\
&- &  2 (g_{tt} + g_{t \phi} \Omega)(g_{t \phi} + g_{\phi \phi}
\Omega)
    \left(\frac{g_{t \phi}}{\rho^2} \right)_{,\alpha \alpha} \nn \\
&+ & \left. \left. (g_{t \phi} + g_{\phi \phi} \Omega)^2
    \left(\frac{g_{t t}}{\rho^2} \right)_{,\alpha \alpha} {} \right] \right) ^{1/2} ,
\eea 
where $\alpha$ stands for $\rho$ (periastron precession), or $z$ (orbital precession). Actually, the
frequencies written above correspond to the difference between the
orbital frequency and the frequency of perturbations in $\rho$, or
$z$, since these differences will show up in
gravitational radiation emitted by the binary as a modulating frequency. 

The energy per unit test-body mass for an equatorial circular orbit
in an axially symmetric spacetime is 
\be 
\label{Eovermu}
\frac{E}{\mu}=\frac{ -g_{t t} - g_{t \phi} \Omega} {\sqrt{-g_{t t}-2
g_{t \phi}\Omega - g_{\phi \phi} \Omega^2} } , 
\ee 
and the
specific energy released as gravitational radiation per logarithmic
interval of frequency is 
\be \label{delE} \frac{\Delta
E}{\mu}=-\Omega \frac{d (E/\mu)}{d \Omega}. 
\ee 
The expression above
assumes that all the energy lost from the test body has been emitted
at infinity as radiation.

The number of gravitational-wave cycles spent in a  logarithmic
interval of frequency, the quantity that could be most accurately measured 
by a gravitational wave detector, is 
\be 
\label{delN} 
\Delta N=\frac{f \Delta
E(f)}{dE_{\textrm{wave}}/dt}, 
\ee 
where $dE_{\textrm{wave}}/dt$ is
the gravitational-wave luminosity, which is assumed to be exactly
the rate of energy loss of the orbiting test body. 
The main contribution of $dE_{\textrm{wave}}/dt$
comes from the mass quadrupole radiative moment of the binary. 
Therefore, when we give the connection between the various  orders of 
$v \equiv (M\Omega)^{1/3}$ (the orbital velocity) in the power series expansion of 
$\Delta N$, and the moments of spacetime, only the highest order moments enter
our expressions, since lower order moments contribute through higher than quadrupole
radiative moments as well in a rather complicated and unclear way. However, at least 
the first 4 power parameters of $\Delta N$ could be written explicitly in terms of the multipole moments.

\section{The metric as a power series of moments}

To get expressions for these 3 measurable quantities, that are
straightforwardly connected to the moments of the central object,
one has to reexpress the
metric functions in terms of all moments, as well as
the radius $\rho$ and the orbital frequency of the test body
$\Omega$.

In our paper, we consider only stationary axisymmetric objects that are reflection symmetric
with respect to their equatorial plane. 
The metric of such a central object alone could be written in $(t,\rho, z, \phi)$ coordinates, in the form of
Papapetrou metric \cite{pap}.
\be
\label{papmet}
ds^2=-F(dt-\omega ~d\phi)^2+\frac{1}{F} \left[
e^{2\gamma} (d\rho^2 + dz^2) + \rho^2 d\phi^2 \right],
\ee
where $F,\omega$, and $\gamma$ are the three functions that fully determine a specific metric. These
are functions of $\rho$ and $|z|$ only, due to axisymmetry and reflection symmetry. Einstein's equations in vacuum
guarantee that once $F$ and $\omega$ are given, $\gamma$ can be easily computed (see \cite{Wald}).
Once we incorporate an electromagnetic field in the vacuum around the compact object, the source
of which is the compact object itself,
which allows spacetime to have the same symmetries, the metric above still describes the
electrovacuum  spacetime, but now the metric and the electromagnetic field should satisfy the
Einstein-Maxwell equations. In order to fully compute the
metric functions one more complex function, $\Phi$,
which is related to the electromagnetic field, is necessary.
$F$, $\omega$, and $\Phi$ themselves can be determined by solving the so-called Ernst equations
\cite{ernst1,ernst2}, which are  the Einstein-Maxwell equations written in a different form.
It is a system of non-linear complex differential equations of second order:
\bea
\label{EinstE}
(\Re ({\mathcal E}) + |\Phi|^2 ) \nabla^2 {\mathcal E} &=&
(\bn {\mathcal E} + 2 \Phi^\ast \bn \Phi )
\cdot \bn {\mathcal E}, \\
\label{EinstF}
(\Re ({\mathcal E}) + |\Phi|^2 ) \nabla^2 {\Phi} &=&
(\bn {\mathcal E} + 2 \Phi^\ast \bn \Phi )
\cdot \bn {\Phi},
\eea
where $\bn$ denotes the gradient in a cartesian 3D space $(\rho,z,\phi)$ and $\Re(\ldots),\Im(\ldots)$, here
and henceforth, denote the real and imaginary
part, respectively, of the complex function in parentheses, while
an asterisc $^\ast$ denotes complex conjugate.
The third metric function, $\gamma$, is then easily computed by integrating the partial derivatives
$\partial \gamma/\partial \rho$, $\partial \gamma/\partial z$, which are given as functions of
derivatives of all other functions (see \cite{ApostSoti}).

From the two complex functions, $\mathcal E$ and $\Phi$, we could construct the metric functions as
\bea
\label{ernstpot}
\mathcal{E}&=&(F-\left|\Phi\right|^2)+i\varphi,
\eea
where $\varphi$ is related with $g_{t\phi}$ through
\bea
\label{intgtf}
g_{t\phi}=F \omega&=& F \int_{\rho}^{\infty} \!\!d\rho'
\frac{\rho'}{F^{2}}\left(\frac{\partial\varphi}{\partial z}+
2{\Re}(\Phi)\frac{\partial{\Im}(\Phi)}{\partial z}-{}\right.\nn\\
& &{}-\left.\left.
2{\Im}(\Phi)\frac{\partial{\Re}(\Phi)}{\partial z}
\right)\right|_{z=\textrm{const}}.
\eea
Note that there is a sign difference in Eq.~(22) of \cite{Ryan}, which has been corrected in
a later paper of Ryan \cite{RyanCORRECT}, and comes from an odd convention of $\omega$
used by Ernst (see relevant comment of \cite{israel}).

Now, instead of $\mathcal E$ and $\Phi$, one could use two new complex functions $\tilde \xi$ and $\tilde q$,
that play the role of gravitational potential and Coulomb potential respectively, and are more directly
connected to the mass and electromagnetic moments of the central body. These potentials are related
to the Ernst functions by
\bea
\label{etoxi}
\mathcal{E} &=& \frac{\sqrt{\rho^{2}+z^{2}}-\tilde{\xi}}{\sqrt{\rho^{2}+z^{2}}+\tilde{\xi}} \\
\Phi&=&\frac{\tilde{q}}{\sqrt{\rho^{2}+z^{2}}+\tilde{\xi}},
\eea
and can be written as power series expansions at infinity
\bea
\label{expxiq}
\tilde {\xi } = \sum\limits_{i,j = 0}^\infty {a_{ij}\bar {\rho }^i\bar {z}^j},\qquad
\tilde {q} = \sum\limits_{i,j = 0}^\infty {b_{ij}\bar {\rho }^i\bar {z}^j},
\eea
where
\be
\label{barred}
\bar \rho \equiv \frac{\rho}{\rho^2+z^2},\qquad \bar z \equiv \frac{z}{\rho^2+z^2},
\ee
and $a_{ij},b_{ij}$ are coefficients that vanish when $i$ is odd.
This reflects the analyticity of the potentials on the $z$-axis. The tilded quantities, here and henceforth,
are the conformally transformed ones, which are essential for calculating the moments (see \cite{Geroch}).

Due to Ernst equations (\ref{EinstE},\ref{EinstF})
the above power expansion coefficients, $a_{ij}$ and $b_{ij}$, are interrelated through 
 complicated recursive relations (see \cite{ApostSoti}).
Essentially, these relations are simply an algebraic version of Einstein-Maxwell equations
for the coefficients of the power expansion of the metric and the electromagnetic field tensor.
The recursive relations could be used to build
the whole power series of $\tilde{\xi}$
and $\tilde{q}$ from a full knowledge of the metric on the axis of symmetry
\bea
\tilde{\xi}(\bar\rho=0)=\sum_{i=0}^{\infty} m_i {\bar z}^i ,\qquad
\tilde{q}(\bar\rho=0)=\sum_{i=0}^{\infty} q_i {\bar z}^i.
\eea

In \cite{SotiApos} a method of calculating the complex multipole moments of the central object in terms of
the $m_{i}$'s and $q_{i}$'s is presented. In brief, the gravitational moments are given by
\begin{equation}
\label{pnfinal}
P_n = \frac{1}{(2n - 1)!!}S_0^{(n)},
\end{equation}
where $S_{a}^{(n)}$ are given by recursive formulas analyzed in \cite{ApostSoti}.

The mass moments $M_{n}$ and the mass-current moments $S_{n}$ are related to $P_{n}$ by
\be
\label{massangular}
P_{n}=M_{n}+i S_{n},
\ee
whereas the electric  moments $E_{n}$ and the magnetic  moments $H_{n}$ are related to $Q_{n}$ by
\be
\label{electricmagnetic}
Q_{n}=E_{n}+i H_{n}.
\ee

Since this algorithm can be used to evaluate the moments in terms of the $m_{i}$'s and $q_{i}$'s,
one can invert these relations and express the $m_{i}$'s and $q_{i}$'s in terms of the moments:
\bea
\label{lom}
m_{n}&=&a_{0n}=M_{n}+iS_{n}+\textrm{LOM},\nn\\
q_{n}&=&b_{0n}=E_{n}+iH_{n}+\textrm{LOM},
\eea
where ``LOM'' stands for lower order multipole moments of any type.
Thus, we can use the recursive relations mentioned above
  to evaluate the $a_{ij}$ and $b_{ij}$ coefficients in terms of the moments.
 
Finally, following the procedure
presented in the beginning of this section 
we can express the metric functions, and their first and second derivatives
as power series of $\rho$ and $z$ with coefficients that are simple algebraic functions of the moments
of the massive body. Since in our study we have confined the motion of the test particle
on the equatorial plane, we  actually need to compute everything at $z=0$ which makes
calculations far simpler than what they seem.

One could argument (see Sec.~IIC of \cite{ApostSoti}) that the reflection symmetry assumed for
the metric is consistent with a set of even electric and odd magnetic moments or
odd electric and even magnetic moments.

\section{Expressions relating the measurable quantities with moments}

Combining the formulae that are given in Sec.~2  we can express the three 
measurable quantities as power series of $v \equiv (M \Omega)^{1/3}$ with coefficients that
have explicit dependence on all four types of moments. The choice of $v$ as a dimensionless parameter
to expand all physical quantities is warranted from the fact that the inspiral phase of a binary,
the best exploitable part in gravitational-wave analysis \cite{CutlFlan}, involves
comparatively low magnitudes of $v$.
All measurable quantities have been transformed to a dimensionless form as well, for example by dividing
the two frequencies $\Omega_\rho$, $\Omega_z$, by the orbital frequency $\Omega$.

We need also a
power series expansion of $\rho$ with respect to $v$, or equivalently $\Omega$. Thus we have to
invert the function $\Omega(\rho)$, at least as a power expansion. From an elementary analysis
of circular geodesics on the equatorial plane (see \cite{Ryan}) we know that
\be
\label{Omega}
\Omega=\frac{ -g_{t \phi,\rho} +
\sqrt{(g_{t \phi,\rho})^2-(g_{tt,\rho})(g_{\phi \phi,\rho})} }{ g_{\phi \phi,\rho} } .
\ee

In the following part of this section we explain the algorithm that one should follow,
in order to obtain the power series for
$\Omega_\rho/\Omega$, $\Omega_z/\Omega$, and $\Delta N$. One starts with a power series of
$\tilde{\xi}$ and $\tilde{q}$ of the form given by Eq.~(\ref{expxiq}). Since no higher than second
derivatives of the metric functions with respect to $z$ are necessary, one should keep $a_{ij}$'s
and $b_{ij}$'s with $0 \leq j \leq 2$, and as many values of $i$ as one needs to carry the power
series expansion of the measurable quantities at a desirable order. All quantities
that are expressed as power series of $\bar{\rho}$ and $\bar{z}$, are evaluated at ${\bar z}=z=0$
at the end, and thus, all expressions are finally power series of ${\bar \rho}=1/\rho$,
due to Eq.~(\ref{barred}). Although
the $a_{ij}$ and $b_{ij}$ are polynomials of various moments, from the practical point of view
it is preferable to keep them as they are, and replace them by their moments
dependence only at the final expressions.
Then from $\tilde{\xi}$ and $\tilde{q}$ we construct ${\mathcal E}$, $\Phi$, and $F$, $\varphi$
(cf., Eqs.~(\ref{ernstpot},\ref{etoxi})). These are sufficient to build all metric functions. 
Next, following the procedure described above,
we expand $\Omega$ as a power series of $1/\rho$, by virtue of Eq.~(\ref{Omega}). This series
is inverted and in this way we obtain $1/\rho$ as a power series of $\Omega$, which
then can easily be turned into a power series of the dimensionless parameter $v$.

Now, the power series representing $1/\rho$ will replace all $1/\rho$ terms appearing at
the expansions of the metric, its derivatives, and all other physical quantities depending on them.  
Finally, one has to rewrite
the $a_{ij}$ and $b_{ij}$ terms appearing at the coefficients of all these power series
as polynomials of the various moments. The recursive relations that
relate all $a_{ij}$ and $b_{ij}$ with  $m_k \equiv a_{0k}$ and $q_k \equiv b_{0k}$,
which are directly related to the scalar moments of spacetime through Eqs.~(24,25) of Ref.~\cite{SotiApos}.

The algorithm described in the previous two paragraphs has been carried out with Mathematica,
and has been checked for the following two subcases: (i) When all electromagnetic fields are
turned off, by erasing all electromagnetic moments ($E_l=H_l=0$), our expressions for
$\Omega_\rho$, $\Omega_z$, $\Delta N$ are identical to the ones computed by
Ryan \cite{Ryan}. (ii) For the Kerr-Newman metric it is quite easy to compute $\Omega$,
$\Omega_\rho$, $\Omega_z$, and $\Delta E/\mu$ for a quasi-equatorial, quasi-circular orbit.
The expressions we  obtain are identical to the ones obtained directly
from the Kerr-Newman metric.

The power series expansion for $\Omega_\rho$, $\Omega_z$, take the following form:
\bea
\label{wpprosw}
\frac{\Omega_{\rho}}{\Omega}&=&\sum\limits_{n=2}^{\infty}R_{n}v^{n},\\
\label{wzprosw}
\frac{\Omega_{z}}{\Omega}&=&\sum\limits_{n=3}^{\infty}Z_{n}v^{n},
\eea
while the corresponding
coefficients, up to 5th order for the two frequencies, for the two
distinct electromagnetic cases ($es$) and ($ms$) are

\bea
\label{wpprosw1}
R^{(es)}_{2}&=&\left(3-\frac{1}{2}\frac{E^{2}}{M^{2}}\right),\quad
R^{(es)}_{3}=-4\frac{S_{1}}{M^{2}},\nn\\
R^{(es)}_{4}&=&\frac{9}{2}-\frac{3}{2}\frac{M_{2}}{M^{3}}-2\frac{E^{2}}{M^{2}}-\frac{13}{24}\frac{E^{4}}{M^{4}},\quad
R^{(es)}_{5}=-10\frac{S_{1}}{M^{2}}-\frac{10}{3}\frac{S_{1}E^{2}}{M^{4}}+5\frac{H_{1}E}{M^{3}},
\eea
\bea
\label{wpprosw2}
R^{(ms)}_{2}&=&\left(3-\frac{1}{2}\frac{H^{2}}{M^{2}}\right),\quad
R^{(ms)}_{3}=-4\frac{S_{1}}{M^{2}},\nn \\
R^{(ms)}_{4}&=&\frac{9}{2}-\frac{3}{2}\frac{M_{2}}{M^{3}}-2\frac{H^{2}}{M^{2}}-
\frac{13}{24}\frac{H^{4}}{M^{4}},\quad
R^{(ms)}_{5}=-10\frac{S_{1}}{M^{2}}-\frac{10}{3}\frac{S_{1}H^{2}}{M^{4}}-5\frac{E_{1}H}{M^{3}},
\eea
\bea
\label{wzprosw1}
Z^{(es)}_{3}&=&2\frac{S_{1}}{M^{2}},\quad
Z^{(es)}_{4}=\frac{3}{2}\frac{M_{2}}{M^{3}},\quad
Z^{(es)}_{5}=-\frac{H_{1}E}{M^{3}}+2\frac{S_{1}E^{2}}{M^{4}},
\eea
\bea
\label{wzprosw2}
Z^{(ms)}_{3}&=&2\frac{S_{1}}{M^{2}},\quad
Z^{(ms)}_{4}=\frac{3}{2}\frac{M_{2}}{M^{3}},\quad
Z^{(ms)}_{5}=\frac{E_{1}H}{M^{3}}+2\frac{S_{1}H^{2}}{M^{4}},
\eea

The fact that we have two new sets of moments (the electromagnetic ones) with respect to Ryan
allows many more combinations of moments in high order terms. Actually, from a practical point of view
these expansions are far more advanced than what will be used in gravitational wave data analysis
in the near future. On the other hand the expressions above present an important feature: in every new order
term a new moment shows up. This suggests that a very accurate observational estimation of the series
could in principle reveal any moment.

A glance at the corresponding terms of the two electromagnetic cases shows that each combination of
moments for the ($es$) case is numerically equal to the corresponding combination
for the ($ms$) case, if the electric and magnetic moments are interchanged.
The sign though is the same for combinations of pure electric, or pure  magnetic moments, but
opposite for combinations of electric and magnetic moments.

Finally, in order to express $\Delta N$ also as a power series of $v$,
we need to expand $dE_{\textrm{wave}}/dt$ as power series of $v$. As was explained in Sec.~2
we cannot work out the perturbative analysis of gravitational wave emission at a generic
spacetime background; we can only obtain accurate expressions for $dE_{\textrm{wave}}/dt$
up to $v^4$ after the leading order. Therefore,
in the following formulae for $\rho$ and $dE_{\textrm{wave}}/dt$, we write simply the power series coefficients
explicitly up to the fourth order.
More specifically, in order to compute the power expansion of $dE_{\textrm{wave}}/dt$ we add up all 
power series contributions (quadrupole radiative moment, current quadrupole radiative moment, and
Post-Newtonian corrections) which are the only ones that show up up to the 4th order. Thus, we yield
\be
\label{powserN}
\Delta N=\frac{5}{96 \pi} \left( \frac{M}{\mu} \right) v^{-5}
\left( 1 + \sum\limits_{n=2}^{\infty} N_{n} v^{n} \right),
\ee
where the $N_n$ coefficients for the two electromagnetic cases are given by the following
polynomials of the moments
\bea
\label{Nfinal}
N^{(es)}_2&=&\frac{743}{336}+\frac{14}{3}\frac{E^{2}}{M^{2}}\nn \\
N^{(es)}_3&=&-4\pi+\frac{113}{12}\frac{S_{1}}{M^{2}}\nn \\
N^{(es)}_4&=&\frac{3058673}{1016064}-\frac{1}{16}\frac{S_{1}^{2}}{M^{4}}
+5\frac{M_{2}}{M^{3}}+{}\nn\\
& &{}+\frac{12431}{504}\frac{E^{2}}{M^{2}}+\frac{179}{9}\frac{E^{4}}{M^{4}}
\eea
\bea
\label{Nfinal2}
N^{(ms)}_2&=&\frac{743}{336}+\frac{14}{3}\frac{H^{2}}{M^{2}}\nn \\
N^{(ms)}_3&=&-4\pi+\frac{113}{12}\frac{S_{1}}{M^{2}}\nn \\
N^{(ms)}_4&=&\frac{3058673}{1016064}-\frac{1}{16}\frac{S_{1}^{2}}{M^{4}}
+5\frac{M_{2}}{M^{3}}+{}\nn\\
& &{}+\frac{12431}{504}\frac{H^{2}}{M^{2}}+\frac{179}{9}\frac{H^{4}}{M^{4}}.
\eea

As in the rest three measurable quantities, the power expansion of $\Delta N$
is such that in every order term a new moment, which was not present in any
lower order term, occurs. This is an indication that all moments can in principle be unambiguously
extracted from accurate measurements of $\Delta N$.

\section{Comments}

As is shown in \cite{Ryan2} the first generation of LIGO is not expected to be able to
extract the first two moments ($S_1$ and $M_2$) with high accuracy ($\sim 0.05$ for the former
and $\sim 0.5$ for the latter one), by analyzing the phase of the waves.
If we allow for electromagnetic fields as well,
the corresponding monopole (which classically is expected to be very close to zero)
will be measured with even higher accuracy than the other two mass moments,
since the charge of the source (or the magnetic monopole in case of some exotic body)
is present at even lower order, namely in the $v^2$ term, while the electric dipole,
or the magnetic dipole, that first show up at the $v^5$ term will be measured with
rather disappointing accuracy. On the other hand, analyzing the data of LISA
leads to accuracies almost two orders of magnitude higher than the corresponding for LIGO.
Thus, it seems quite promising that LISA will give us the opportunity to measure the first few
moments, including the electromagnetic dipole moments, quite accurately.
Also, the fact that in every new order in the power series of $\Delta N$
a new moment appears is significant, since this means that in principle a unique set of moments
arises from an accurate estimation of all power series terms. Actually, there are two possible sets of
moments; one for each electromagnetic case, since we cannot \textit{a priori} exclude one of them.
We can only exclude one of the two sets on physical grounds, if only one of them leads to
a physically reasonable classical object (for example a highly magnetized compact object
is physically preferable to a compact object with a huge electric dipole). If we manage
to measure a few lower moments, we can check if they are interrelated as in a Kerr-Newman
metric \cite{SotiApos}. A positive outcome of such a test will be of support to the black-hole no-hair theorem
in the case that the central object is a black-hole. The case of observational violation of
the black-hole no-hair theorem could either mean that the central object is not a black-hole,
or that the theorem does not hold. Of course to assume the latter an extra verification that the central object is
indeed a black hole is necessary. 

While the phase of a gravitational wave is the quantity that can be most accurately
measured, since a large number of cycles (a few thousand for LIGO and a few hundred thousand
for LISA in case of binaries with high-ratio of masses) is sweeping up the sensitive part of the detectors,
the two precession frequencies $\Omega_\rho$ and $\Omega_z$, can in principle be measured
if the detectors become more sensitive and templates that describe modulating waves are used \cite{Apos}.
If this ever become possible one could use any of them to test the no-hair theorem. This
would demand no more than the four lower order terms, since according to this theorem all moments
depend on only three quantities (mass, angular momentum, and total charge).
Actually, from measurements of modulating frequencies we could not at first determine
which frequency corresponds to each precession. However, the power expansions of the two
frequencies begin at a different order, and thus we could discern them.
Unfortunately, there are terms in these expansions that contain more than one first occurring moment
(for example the first term of $\Omega_\rho$, $R_2$, is a function of $E$ or $H$ and $M$).
However, expansions of $\Omega_\rho$ and $\Omega_z$, if used simultaneously, along with
the intrinsic dependence of $v$ on $M$, could finally lead to the full determination of the moments.
Notice though that some corresponding terms in the series expansions of the two frequencies
depend on the same set of multipole moments, like $R_3$ and $Z_3$. Assuming that the observed system
is sufficiently well described by our model for the binary, this multipole information could serve
to determine these moments with better precision.

If we manage to obtain information about the precessing frequencies
or the evolution of the orbits of small bodies around compact stars by other observational means, we could
measure the multipole moments (mass moments, mass-current moments, and electromagnetic moments)
of the compact central object and thus put strict restrictions on the models that are used to
describe the interior of these objects.

Our analysis demonstrates that it will be possible to determine
all types of multipole moments of the central object, from future gravitational wave measurements.
Thus, apart from  spacetime geometry, we could also determine the central body's
electromagnetic fields.
Although the data of LISA should be  suitable for extracting such information with high accuracy,
the assumptions of circular and equatorial orbit are not that
realistic. From this point of view we consider our work as a step towards a more detailed
analysis with not so restrictive assumptions.


\begin{theacknowledgments}
 This research was supported in part by Grant No 70/4/4056 of the Special Account for Research Grants of the
University of Athens, and in part by Grant No 70/3/7396 of the ``PYTHAGORAS'' research funding program.

\end{theacknowledgments}



\bibliographystyle{aipprocl} 


\begin{thebibliography}{9}

\bibitem{Geroch} R.~Geroch, J.~Math.~Phys. {\bf 11}, 2580 (1970).
\bibitem{Hansen} R.~O.~Hansen, J.~Math.~Phys. {\bf 15}, 46 (1974).
\bibitem{Simon1} W.~Simon and R.~Beig, J.~Math.~Phys. {\bf 24}, 1163 (1983).
\bibitem{Simon2} W.~Simon, J.~Math.~Phys. {\bf 25}, 1035 (1984).
\bibitem{Ryan} F.~D.~Ryan, Phys.~Rev.~D {\bf 52}, 5707 (1995).
\bibitem{pap} A.~Papapetrou, Ann.~Phys. {\bf 12}, 309 (1953).
\bibitem{Wald} R.~M.~Wald, {\it General Relativity} (The University of Chicago Press, Chicago, 1984).
\bibitem{ernst1} F.~J.~Ernst, Phys.~Rev. {\bf 167}, 1175 (1968).
\bibitem{ernst2} F.~J.~Ernst, Phys.~Rev. {\bf 168}, 1415 (1968).
\bibitem{ApostSoti} T.~P.~Sotiriou and T.~A.~Apostolatos, Phys.~Rev.~D {\bf 71}, 044005 (2005).
\bibitem{RyanCORRECT} See Ref. [16] of F.~D.~Ryan, Phys.~Rev.~D {\bf 56}, 7732 (1997) and Eq.~(2.10) of M.~Shibata and M.~Sasaki,
Phys.~Rev.~D {\bf 58}, 104011 (1998).
\bibitem{israel} W.~Israel, Phys.~Rev.~D {\bf 2}, 641 (1970).
\bibitem{SotiApos} T.~P.~Sotiriou and T.~A.~Apostolatos, Class.~Quantum Grav. {\bf 21}, 5727 (2004).
\bibitem{CutlFlan} C.~Cutler and \'E.~E.~Flanagan, Phys.~Rev.~D {\bf 49}, 2658 (1994).
\bibitem{Ryan2} F.~D.~Ryan, Phys.~Rev.~D {\bf 56}, 1845 (1997).
\bibitem{Apos} T.~A.~Apostolatos Phys.~Rev.~D {\bf 54}, 2421 (1996).
\end{thebibliography}

%


\end{document}